\documentstyle[11pt,psfig]{article}
\def\reference{\parskip 0pt\par\noindent\hangindent 0.5 truecm}
\def\kms{km ${\rm s}^{-1}$}
\def\kms{\hbox{$\rm\thinspace km s^{-1}$}}
\def\kmsMpc{\hbox{$\rm\thinspace km s^{-1} Mpc^{-1}$}}
\def\hMpc{\hbox{$\rm\thinspace h^{-1}_{75}Mpc$}}
\def\mo{\hbox{${\rm\thinspace M}_{\odot}$}}
\begin{document}
%
%
\title{New Structure In The Shapley Supercluster}
%


\author{M.J.Drinkwater$^{1}$ \and
D.Proust$^{2}$ \and
Q.A.Parker$^{3}$ \and
H.Quintana$^{4}$ \and
E.Slezak$^{5}$
} 

\date{Accepted for {\em Publications of the Astronomical
Society of Australia} 1999 Feb 22}
\maketitle

{\center
$^1$ School of Physics, University of New South Wales, Sydney 2052,
Australia, mjd@phys.unsw.edu.au\\[3mm]
$^2$ DAEC, Observatoire de Meudon, 92195 Meudon Cedex,
France, proust@obspm.fr\\[3mm]
$^3$ Anglo-Australian Observatory, Coonabarabran, NSW 2357, Australia,
qap@aaocbn.aao.gov.au\\[3mm]
$^4$ Departamento de Astronomia y Astrofisica,
Pontificia Universidad Cat\'olica de Chile\\
Casilla 104, Santiago 22, Chile,
hquintana@astro.puc.cl\\[3mm]
$^5$ Observatoire de Nice, 06304 Nice Cedex 4 France,
slezak@obs-nice.fr\\[3mm]
}

%
\begin{abstract}
We present new radial velocities for 306 bright ($R<16$) galaxies in a 77
deg$^2$ region of the Shapley supercluster measured with the
FLAIR-II spectrograph on the UK Schmidt Telescope.

The galaxies we measured were uniformly distributed over the survey
area in contrast to previous samples which were concentrated in
several rich Abell clusters. Most of the galaxies (230) were members
of the Shapley supercluster: they trace out two previously unknown
sheets of galaxies linking the Abell clusters of the supercluster. In
a 44 deg$^2$ area of the supercluster excluding the Abell clusters,
these sheets alone represent an overdensity of a factor of $2.0\pm0.2$
compared to a uniform galaxy distribution.

The supercluster is not flattened in the Declination direction as was
suggested in previous papers. Within our survey area the new galaxies
contribute an additional 50\% to the known contents of the Shapley
supercluster with a corresponding increase in its contribution to the
motion of the local group.
\end{abstract}


\bigskip

%
%

\section{Introduction}

The Shapley supercluster (SSC) has been investigated by numerous authors
since its discovery in 1930 (Quintana et al.\ 1995; hereafter Paper
I). It lies in the general direction of the dipole anisotropy of the
Cosmic Microwave Background (CMB), and is located at 130\hMpc\ beyond
the Hydra-Centaurus supercluster ($\simeq 50$\hMpc\ away from us). It
consists of many clusters and groups of galaxies in the redshift range
$0.04<z<0.055$.  The central cluster A3558 has also been measured with
a ROSAT PSPC observation by Bardelli et al.\ (1996) who derive a total
mass of $M_{tot}=3.1\times10^{14}\mo$ within an Abell radius of
2\hMpc. Several other x-ray clusters form part of the Shapley
supercluster (Pierre et al.\ 1994).

The Shapley supercluster is recognised as one of the most massive
concentrations of galaxies in the local universe (Scaramella et al.\
1989; Raychaudhury 1989), so it is of particular interest to
consider its effect on the dynamics of the Local Group. In Paper I it
was estimated that for $\Omega_{o}= 0.3$ and $H_{o}= 75\kmsMpc$ the
gravitational pull of the supercluster may account for up to 25\% of
the peculiar velocity of the Local Group required to explain the
dipole anisotropy of the CMB radiation, in which case the mass of the
supercluster would be dominated by inter-cluster dark matter.

Previous studies of the Shapley supercluster (Paper I, Quintana et
al.\ 1997; hereafter Paper II) have concentrated on the various rich
Abell galaxy clusters in the region, but this might give a very biased
view of the supercluster.  As was noted in Paper I, ``the galaxy
distribution inside the supercluster must be confirmed by the
detection in redshift space of bridges or clouds of galaxies
connecting the different clusters''.

We are continuing this project, using data from wide-field multi-fibre
spectrographs to measure many more galaxy redshifts and get a more
complete picture of the composition of the supercluster. Our main aims
are first to define the real topology of the SSC: in Paper I it was
shown that the SSC is significantly flattened, but the real extent of
the concentration is not well defined. Secondly we will analyse the
individual X-ray clusters that are true members of the Shapley
Supercluster in order to estimate the cluster masses, and investigate
suspected sub-structure. Additional observations are planned before we
present a full analysis of the dynamics (Proust et al.\ 1998 in
preparation).

In Paper II we presented data from the MEFOS spectrograph on the
European Southern Observatory 3.60m telescope. This has 30 fibres in a
1 deg diameter field, so the observations were again mainly
concentrated on the known clusters, determining for several of them if
they were members of the supercluster or not.

In this paper we present new data obtained with the FLAIR-II (Parker
\& Watson 1995; Parker 1997) multi-fibre spectrograph on the UK
Schmidt Telescope at the Anglo-Australian Observatory. This has 90
fibres in a $5.5\times5.5$ deg$^2$ field and has allowed us to measure
a much more uniform distribution of galaxies in the direction of the
SSC, avoiding the previous bias in favour of the rich clusters.  Our
data reveal the existence of a sheet of galaxies connecting the main
parts of the supercluster.  We describe the sample and observations in
Section~\ref{sec_obs}.  We present the results along with previous
measurements in Section~\ref{sec_res} and discuss the significance of
the measurements in Section~\ref{sec_discuss}.


\section{Observations}
\label{sec_obs}
\label{sec_sel}

Although a large body of galaxy velocity data is available in the
literature for the SSC, the existing samples of redshifts in each
cluster are highly incomplete, even at the bright end of the
luminosity function. We have therefore started a campaign to obtain
complete samples down to the same magnitude below $L_*$ for each
cluster. Each selected cluster has a projected diameter of 2.5 to 3.0
degrees, so the FLAIR-II system on the UKST with a $5.5\times5.5$
deg$^2$ field is an ideal facility for this project. The very wide
field also permits us to probe the regions between the dominant
clusters neglected in previous observations. In this paper we
emphasise our results from these regions.

We selected targets from red ESO/SRC sky survey plates scanned by the
MAMA machine at Paris Observatory (as described in Paper II; see also
Infante et al.\ 1996). The fields observed (listed in Table~\ref{tab_obs})
were the standard survey fields nearest to the centre of the
cluster (13:25:00 $-$31:00:00 B1950). These covered an
area of 77 deg$^2$ allowing us to probe the limits of the SSC out to
radii as large as 8 deg.

We defined a sample of galaxies to a limit of $R<16$, corresponding
(assuming a mean $B-R=1.5$) to $B<17.5$, the nominal galaxy limiting
magnitude of the FLAIR-II system. This corresponds to an absolute
magnitude of $M_B=-19$ at the Shapley distance of 200\hMpc.  This gave
samples of 600--1000 galaxies per field. We then removed any galaxies
with published measurements in the NED database or measured by
H. Quintana and R. Carrasco (private communication, 1997): 46 galaxies
for F382, 81 for F383 and 200 for F444. For each observing run we then
selected random subsamples of about 110 targets per field from the
unobserved galaxies. When preparing each field for observation at the
telescope we made a further selection of 80 targets to observe (10
fibres being reserved for measurement of the sky background). This
final selection was essentially random, but we did reject any galaxies
too close (less than about 1 arcminute) to another target already
chosen or a bright star.

\begin{table}
\caption{Journal of FLAIR observations}
\label{tab_obs}
\bigskip
\centering
\begin{tabular}{lrcrrrrr}
\hline
Date        & Field & RA (1950) Dec& exposure & seeing & weather & N$_{z}$   \\
\hline
1997 May 5  &  F382 & 13:12:00 $-$35:00:00 & 18000s & 2--3'' & cloud & 69 \\
1997 May 6  &  F444 & 13:25:00 $-$30:00:00 & 15000s & 2--3'' & cloud & 47 \\
1997 May 8  &  F383 & 13:36:00 $-$35:00:00 & 18000s & 3--5'' & clear & 73 \\
1998 April 25& F383 & 13:36:00 $-$35:00:00 & 15000s & 2--3'' & clear & 61 \\
1998 April 27& F382 & 13:12:00 $-$35:00:00 & 21000s & 3--5'' & cloud & 56 \\

total       &       &                      &        &        &       &306\\
\hline
\end{tabular}

\smallskip

{\small Note: N$_{z}$ is the number of galaxies with measured redshifts in each
field.}
\end{table}

We observed a total of 3 fields with the FLAIR-II spectrograph in 1997
May and two more in 1998 April. The details of the observations are
given in Table~\ref{tab_obs}. In 1997, out of 6 allocated nights we
were only able to observe 3 FLAIR fields successfully due to poor
weather and the first of these was repeated over 3 nights. Field F444
was observed in particularly poor weather resulting in a much lower
number of measured redshifts. In 1998 we again had poor weather, and
were only able to observe two fields in an allocation of 8
half-nights.

The data were reduced as in Drinkwater et al.\ (1996) using the
dofibers package in IRAF (Tody 1993).  We measured the radial
velocities with the RVSAO package (Kurtz \& Mink 1998) contributed to
IRAF.

Redshifts were measured for absorption-featured spectra using the
cross-correlation task XCSAO in RVSAO. We decided to adopt as the
absorption velocity the one associated with the minimum error
from the cross-correlation against the templates. In the great
majority of cases, this coincided also with the maximum R parameter of
Tonry \& Davis (1979).  The redshifts for the emission line objects
were determined using the EMSAO task in RVSAO. EMSAO finds emission
lines automatically, computes redshifts for each identified line and
combines them into a single radial velocity with error. Spectra
showing both absorption and emission features were generally measured
with the two tasks XCSAO and EMSAO and the result with the lower error
used. In two spectra with very poor signal (13:05:19.9 $-$33:00:31 and
13:23:22.9 $-$36:47:09) the emission lines were measured manually and
a conservative error of 150\kms\ assigned.  We measured velocities
successfully for 306 galaxies in the sample: these are presented in
Table~\ref{tab_data}.

We have compared the distributions of the galaxies we measured to the
input samples to check they are fair samples. There is no significant
difference in the distributions of the coordinates but there is a
small difference in the magnitude distributions in the sense that the
measured sample does not have as many of the faintest galaxies as the
input sample. This is to be expected as these would have the lowest
signal in the FLAIR-II spectra, but this should not affect our study
of the spatial distribution significantly.

\section{Results}
\label{sec_res}

\begin{figure}
\begin{center}
\vspace{-1.5cm}
\hspace{1.0cm}
\psfig{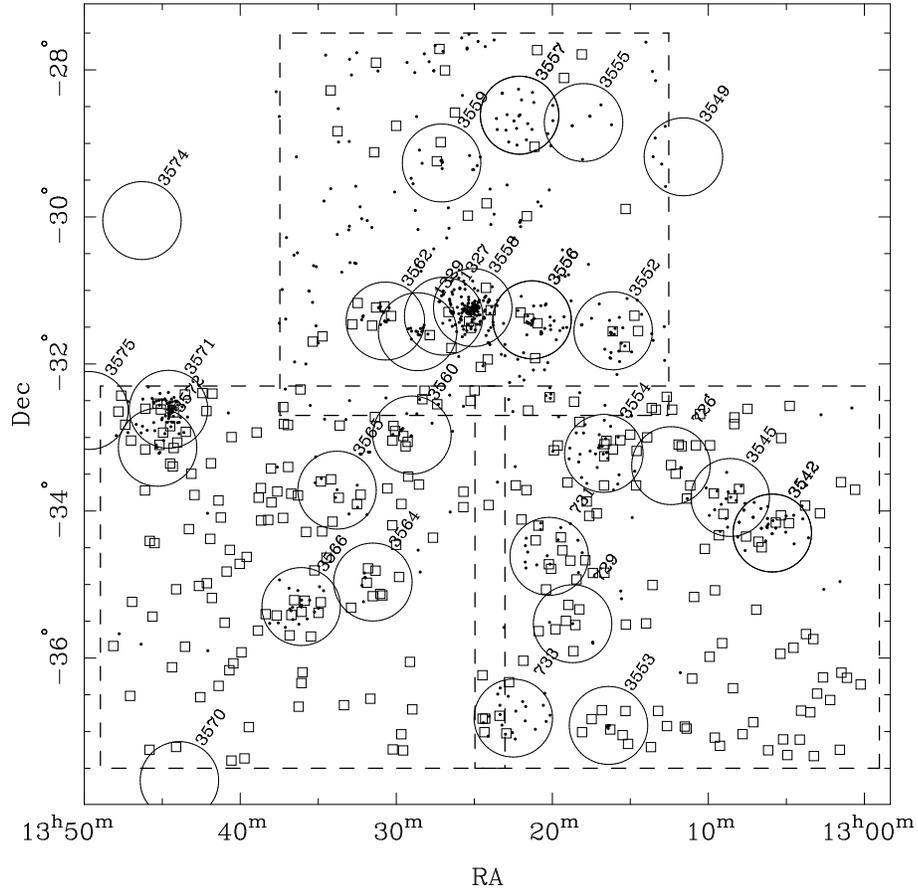}
\caption{Distribution of galaxies measured with FLAIR-II (squares) in
the three UK Schmidt fields (large dashed squares) observed. The
co-ordinate axes are for equinox B1950. Also shown are previously
measured galaxies (dots) and known Abell clusters (labeled circles).}
\label{fig_sky1}
\end{center}
\end{figure}

Previous studies of the SSC have covered a very large region of sky, but
we will limit our analysis in this paper to the region of sky we
observed with FLAIR-II: the three UK Schmidt fields in
Table~\ref{tab_obs}. In some cases we will further restrict our
analysis to the two Southern fields F382 and F383 where our
observations were much more complete. The distribution of these fields
and the galaxies we observed is shown in Fig.~\ref{fig_sky1}. We also
show any previously observed galaxies and the known Abell Clusters.

\begin{figure}
\begin{center}
\vspace{-1cm}
\hbox{
\hspace{-3.5cm}
\psfig{file=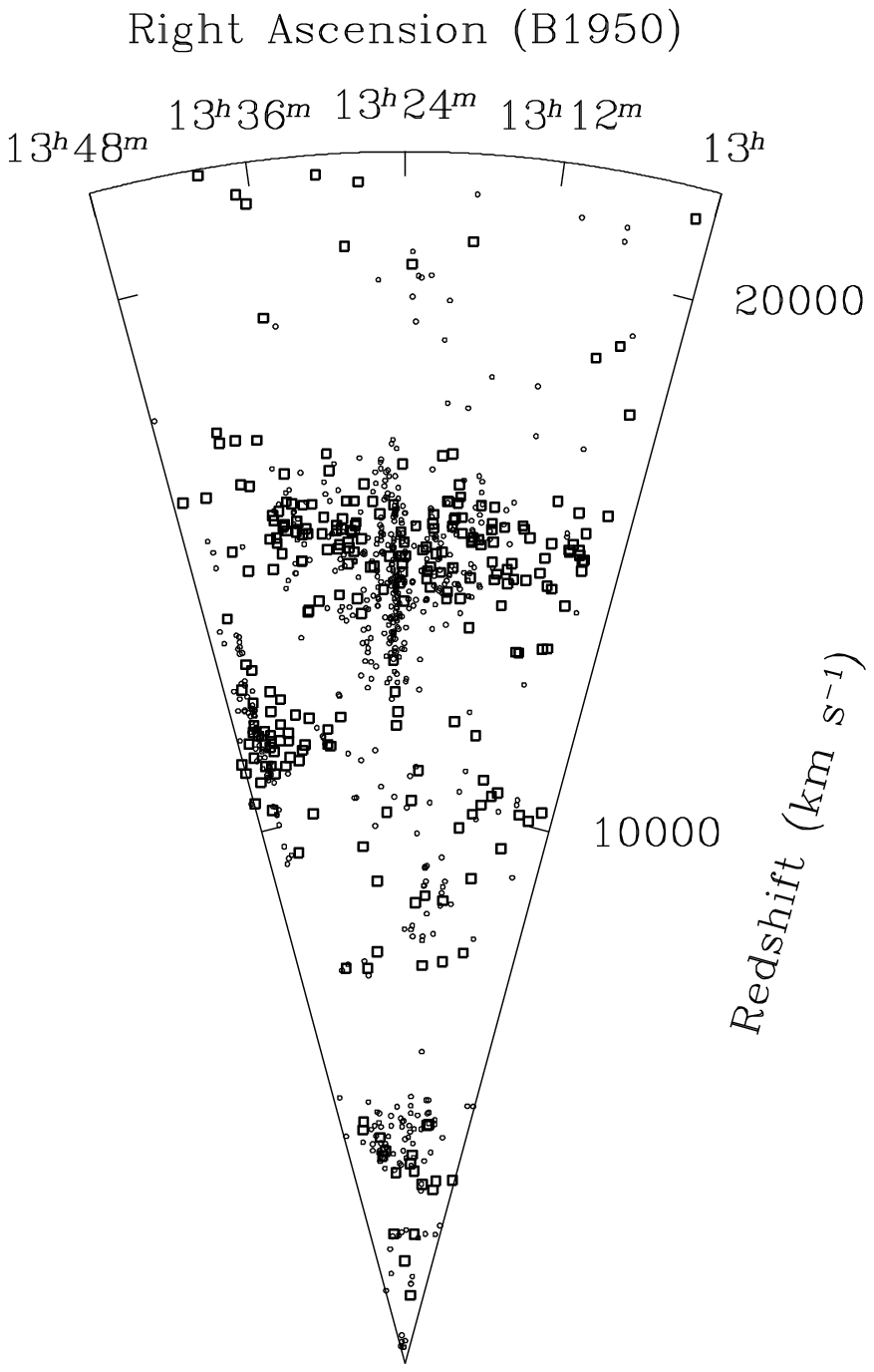,width=14.0cm}
\hspace{-8.9cm}
\psfig{file=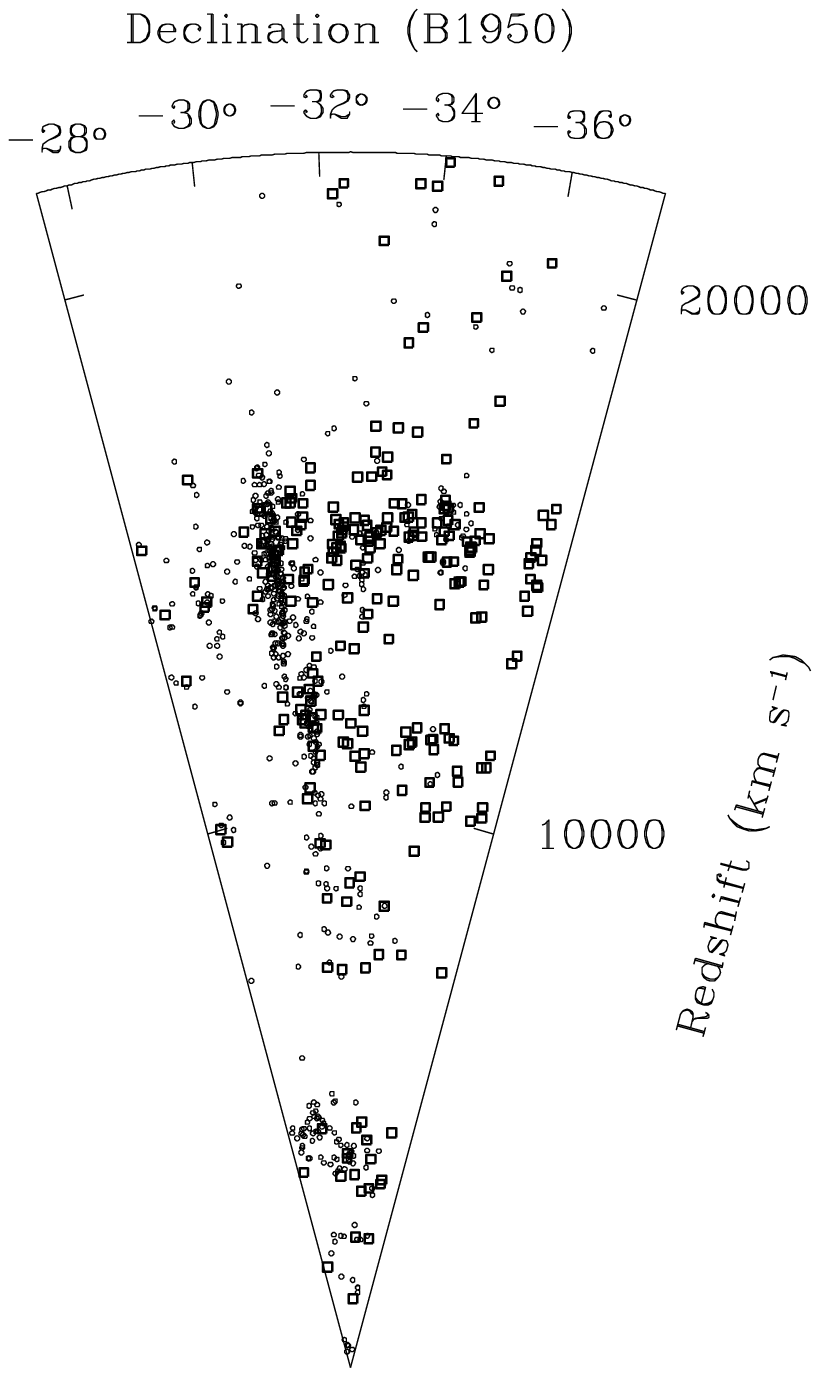,width=14.0cm}
\hspace{-8.9cm}
\psfig{file=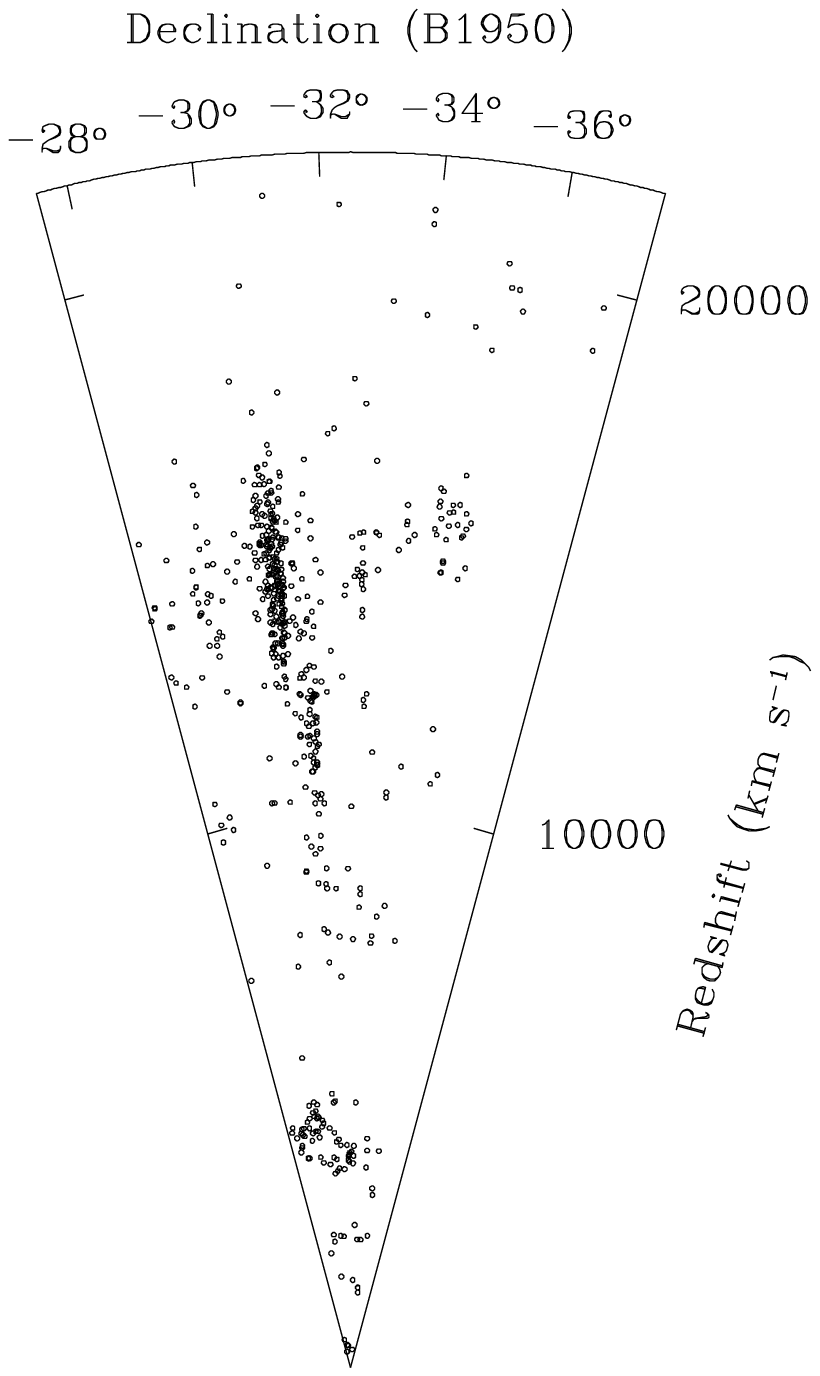,width=14.0cm}
}
\vspace{-1cm}
\caption{Cone diagrams of all known galaxy redshifts in the direction
of the Shapley supercluster. Previously published galaxies are plotted
as dots; the new measurements are plotted as open squares. The angular
scale is enlarged by a factor of 3 for clarity. The distribution in
Declination is repeated at the Right for comparison without the new
measurements.}
\label{fig_cone1}
\end{center}
\end{figure}

\begin{figure}
\begin{center}
\vspace{-2cm}
\hspace{0.7cm}
\psfig{file=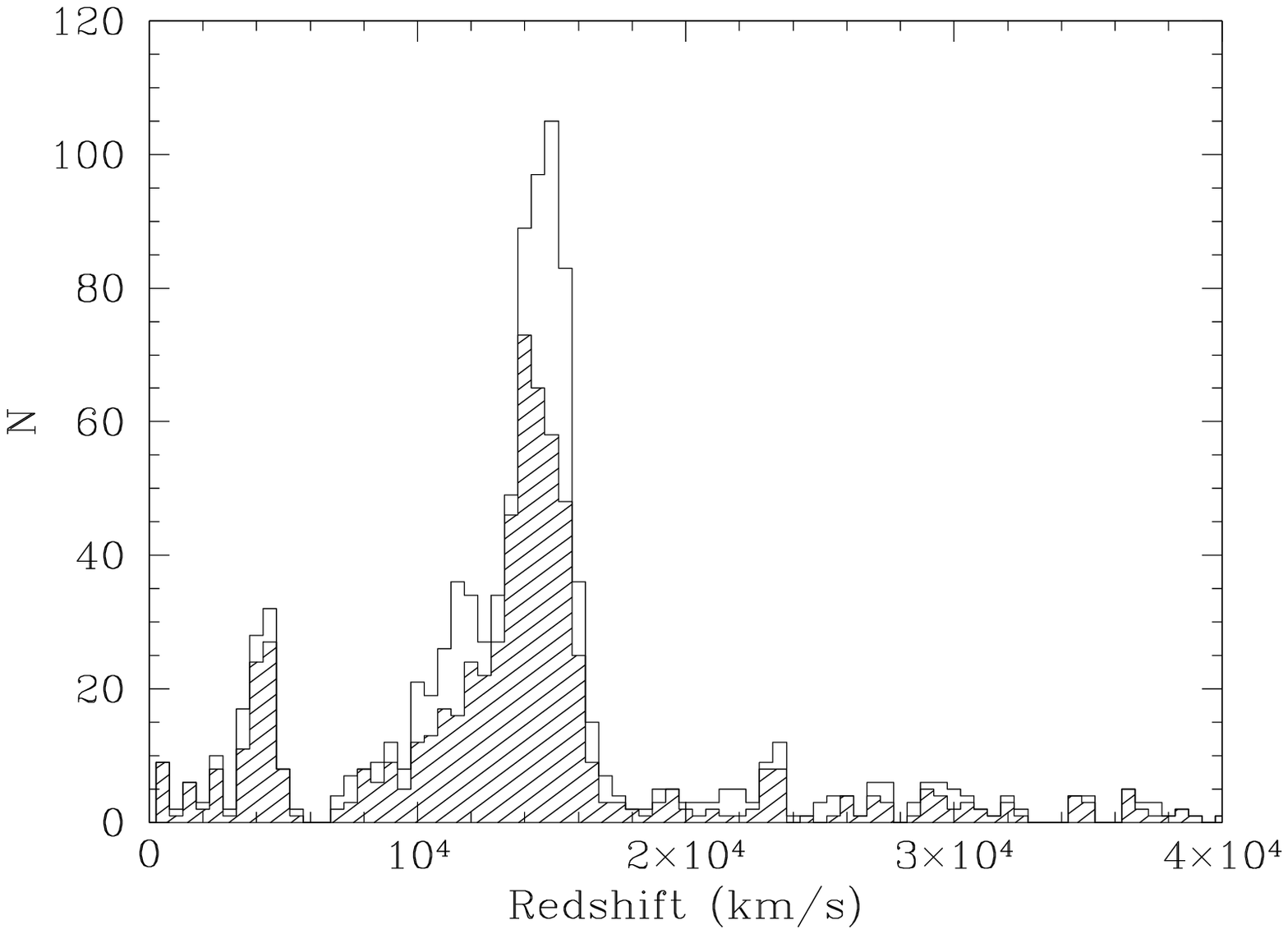,width=14.0cm}
\caption{Histogram of galaxy redshifts in the direction of the Shapley
supercluster in the region defined by the three Schmidt fields shown
in Fig.~\ref{fig_sky1} with a step size of 500\kms. The upper
histogram is for all the data and the lower histogram just gives the
previously published data.}
\label{fig_histo}
\end{center}
\end{figure}

We present the resulting distribution of galaxies towards the Shapley
supercluster in Fig.~\ref{fig_cone1} as cone diagrams
and in Fig.~\ref{fig_histo} as the histogram of
all velocities up to $40000\kms$. The importance of the SSC in this
region of the sky is demonstrated by the fact the fully three quarters
of the galaxies we measured belong to the SSC with velocities in the
range 7580--18300\kms.  In all the plots the new data are indicated
by different symbols to emphasise their impact (this can also be seen
by comparing these figures with the equivalent ones in Paper II).  It
can be seen that by probing large regions of the SSC away from the
rich Abell clusters, we have revealed additional structure which we
discuss in the following sections.

\subsection{Foreground Galaxies}

First in agreement with previous work we also note the presence of a
foreground wall of 269 galaxies (Hydra-Centaurus region) at
$\overline{V}= 4242\kms$ with $\sigma = 890\kms$ in the range
$2000-6000\kms$. This distribution can be related with the nearby
cluster A3627 associated with the ``Great Attractor'' (Kraan-Korteweg
et al.\ 1996).

\subsection{Clusters in the Shapley Supercluster}

\begin{figure}
\begin{center}
\vspace{-1.5cm}
\hspace{1.0cm}
\psfig{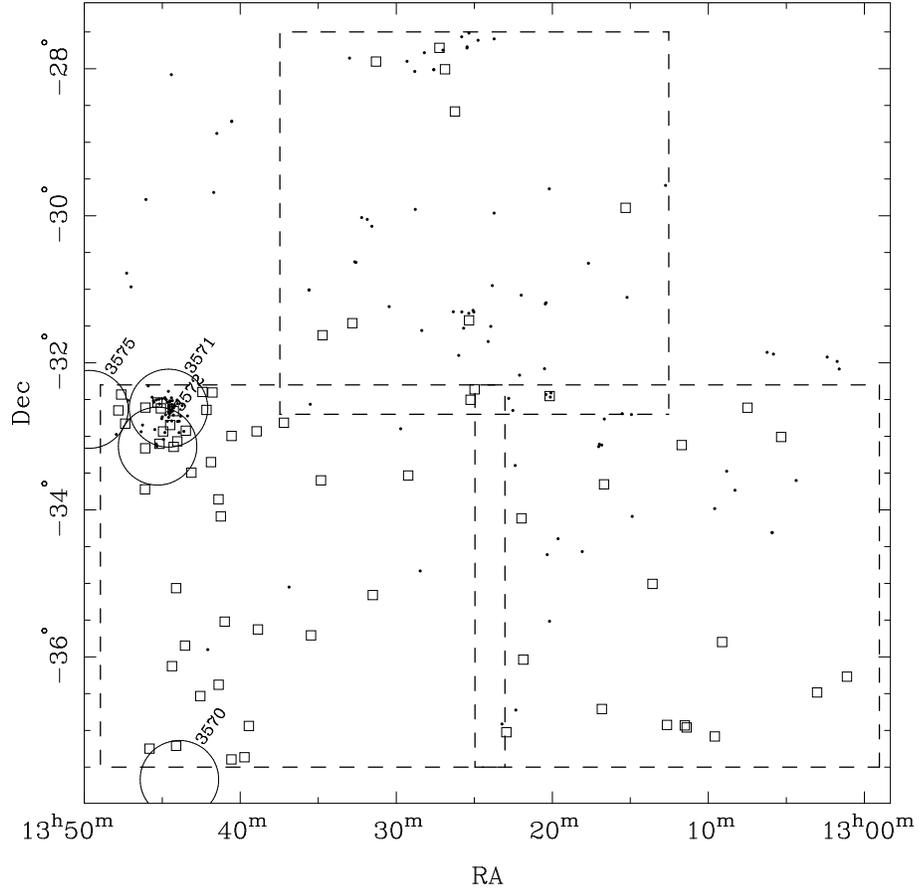}
\caption{Galaxies and clusters in the direction of the Shapley
supercluster with velocities in the range $7580<v<12700\kms$ (near
side of the main supercluster). Previously published galaxies are
plotted as dots and our new measurements are plotted as open squares.
Abell clusters in this velocity range are plotted as labeled circles
of radius 0.5 deg (approximately 1 Abell radius at this distance).
}
\label{fig_skya}
\end{center}
\end{figure}

\begin{figure}
\begin{center}
\vspace{-1.5cm}
\hspace{1.0cm}
\psfig{file=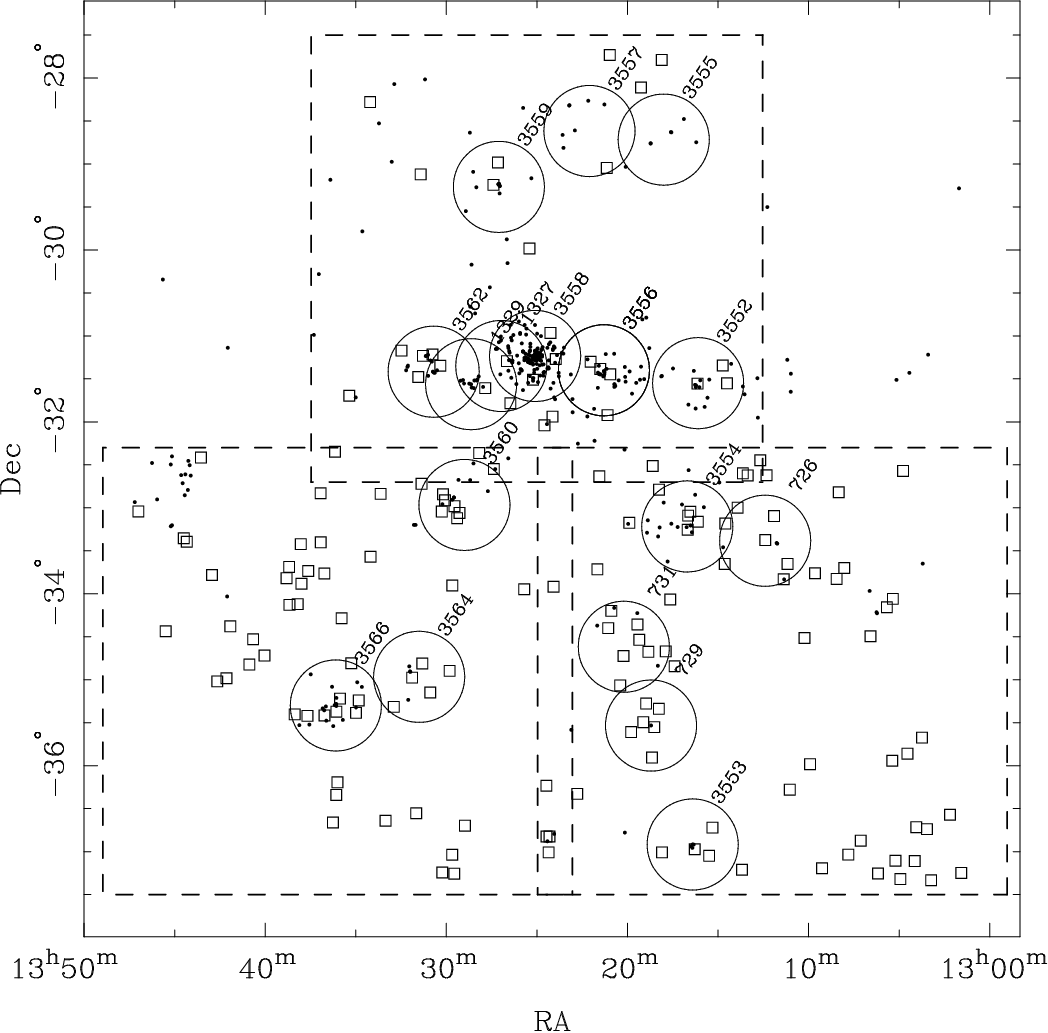,width=12.0cm}
\caption{Galaxies and clusters in the direction of the Shapley
supercluster with velocities in the range $12700<v<18300\kms$
(main supercluster).  Previously published galaxies are
plotted as dots and our new measurements are plotted as open squares.
Abell clusters in this velocity range are plotted as labeled circles
of radius 0.5 deg (approximately 1 Abell radius at this distance).
}
\label{fig_skyb}
\end{center}
\end{figure}

The previous observations reported in Papers I and II concentrated on
the Abell clusters, clarifying the location of many of them. We
reproduce a list of the main clusters in the SSC region in
Table~\ref{tab_clus} for reference and plot their positions in
Fig.~\ref{fig_sky1}.  As noted above, our new measurements concentrate
on galaxies outside the rich clusters in this field. In particular we
observed virtually no galaxies in foreground or background clusters.
We compare the distribution of the SSC galaxies to the Abell clusters
in two velocity slices in Figs.~\ref{fig_skya} and~\ref{fig_skyb}.

In the near side of the SSC ($7580<v<12700\kms$: Fig.~\ref{fig_skya})
we detected several new galaxies in the clusters A3571 and A3572. This
region has a very extended velocity structure with several galaxies in
the higher range (Fig.~\ref{fig_skyb}).  At the velocity of the main
part of the SSC ($12700<v<18300\kms$: Fig.~\ref{fig_skyb}) we have
found additional galaxies in many of the clusters, especially the
poorer ones like AS726, AS731 and A3564. The main conclusion however
is that the clusters are seen as peaks in a sheet-like distribution
rather than isolated objects.

\subsection{Structure of the Shapley Supercluster}

The main impact of our new data is to revise our knowledge of the
large-scale structure of the SSC by measuring a large number of
galaxies away from the rich Abell clusters previously studied.  The
majority of the galaxies we observed were part of the SSC, so our
principal result is to show that the SSC is bigger than previously
thought with an additional 230 galaxies in the velocity range
$7580<v<18300\kms$ compared to 492 previously known in our survey area.

Looking at the cone diagrams (Fig.~\ref{fig_cone1}) and the velocity
histogram in Fig.~\ref{fig_histo} our first new observation is that
the SSC is clearly separated into two components in velocity space,
the nearer one at $\overline{v}=10800\kms$ ($\sigma_v=1300\kms$) to
the East of the main concentration at $\overline{v}=14920\kms$
($\sigma_v=1100\kms$). The two regions contain 200 and 522 galaxies
respectively.  Some evidence for this separation was noted in the
velocity distribution in Paper II, but it is much clearer with our new
data.

Secondly, it can be see from the Declination cone diagram in
Fig.~\ref{fig_cone1} as well as the sky plots in Figs.~\ref{fig_skya}
and~\ref{fig_skyb} that the Southern part of the SSC consists of two
large sheets of galaxies of which the previously measured Abell
Clusters represent the peaks of maximum density. 

\begin{figure}
\begin{center}
\vspace{-2cm}
\hspace{0.7cm}
\psfig{file=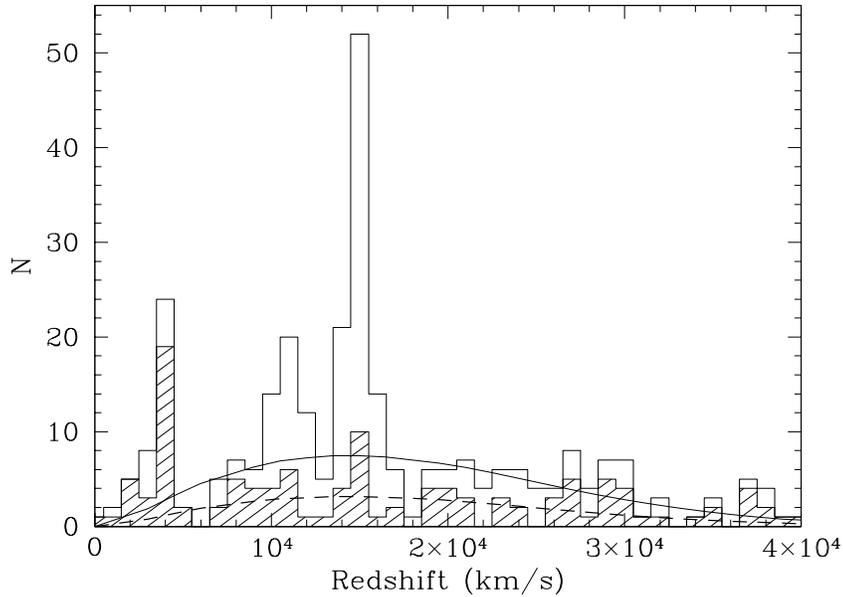,width=14.0cm}
\caption{Histogram of all galaxy redshifts in Southern inter-cluster
region the Shapley supercluster with a bin size of 1000\kms. The upper
histogram is for both our new data and the previously published data 
and the lower histogram just gives the old data. For each histogram a
curve shows the predicted distribution allowing for the size of the
region and the completeness of the samples based on a uniform galaxy
distribution (Metcalfe et al.\ 1991). }
\label{fig_histo2}
\end{center}
\end{figure}

To consider the significance of this extended distribution of galaxies
it is helpful to define an inter-cluster sample consisting of galaxies
in the Southern fields (F382 and F383) outside the known Abell
clusters in the SSC velocity range. We eliminated all galaxies within
a 0.5 degree radius (about 1 Abell radius) of all the clusters shown
in Figs.~\ref{fig_skya} and~\ref{fig_skyb}. Very few of the
previously-measured galaxies remain in the sample. In
Fig.~\ref{fig_histo2} we plot a histogram of the galaxy velocities in
this inter-cluster sample compared to the predicted $n(z)$
distribution of galaxies. The predicted distribution was based on the
number counts of Metcalfe et al.\ (1991) normalised to the area of the
Southern sample after removing clusters (44 deg$^2$) and corrected for
completeness (304 out of a possible 1194 galaxies measured in total).
We also show the histogram (shaded) and predictions (dashed) for the
previously-measured galaxies in the same field (128 out of a possible
1194). The histogram shows that even for the inter-cluster galaxies
there is a large overdensity in the SSC region ($7500<cz<18500$\kms):
we measure 161 galaxies compared to 74 expected. This is an
overdensity of $2.0\pm0.2$ detected at the 10 sigma level. This is averaged
over the whole SSC velocity range; the overdensity in individual
1000\kms\ bins peaks at about 7.  By comparison the previous data (42
galaxies, 33 expected) gave an overdensity of 1.3 detected at only 1.5
sigma. The overdensity for the whole SSC including the Abell clusters
is, of course, much larger still.

These new observations mean that we must modify the conclusions of
Paper I about the overall shape of the SSC. In Paper I it was
concluded from the velocity distribution of the clusters that the SSC
was very elongated and either inclined towards us or rotating.  The
SSC extends as far as our measurements to the South, so we find it is
not elongated or flattened.  We now suggest that it is more complex
still, being composed of the known Abell clusters embedded in two
sheets of galaxies of much larger extent.

\section{Discussion}
\label{sec_discuss}

Our new observations of galaxies towards the Shapely supercluster
have, by surveying a large area away from known clusters, revealed
substantial new large structures in the region. The cluster is part of
a much large structure than was apparent from the previous
observations, extending uniformly in two sheets over the whole region
we surveyed to the South of the core of the SSC.  We detected an
additional 230 members of the SSC in our whole survey area,
representing a 50\% increase on the previous total of 492 SSC
galaxies.  Our measurements to the North of the cluster were much less
complete (only one field in poor weather) so we cannot exclude the
possibility that these sheets of galaxies extend equally to the North.
Recent results by Bardelli, Zucca \& Zamorani (1999) support this
possibility: they have measured galaxies in 18 small (40 arcmin)
inter-cluster fields North of the core of the SSC and also find an
overdensity at the SSC velocity.

In Paper I the effect of the SSC on the dynamics of the Local Group
was estimated. It was found that the mass in the cluster could account
for at least 25\% of the motion of the Local Group with respect to the
cosmic microwave background. Our new data suggest that the SSC is
at least 50\% more massive with a significant part of the extra mass in
the closer sub-region. The SSC therefore has a more important
effect on the Local Group that previously thought, although we
defer a detailed calculation until we have additional data
(Proust et al.\ 1999, in preparation).

\section*{Acknowledgements}

We wish to thank Roberto de Propris for kindly providing the software
to calculate the predicted galaxy $n(z)$ distributions and we are
grateful of the staff of the UKST and AAO for their assistance in the
observations. This research was partially supported by the cooperative
programme ECOS/ CONICYT C96U04 and HC acknowledges support from a
Presidential Chair in Science. MJD acknowledges receipt of an AFCOP
travel grant and a French Embassy Fellowship in support of visits to
Paris Observatory where some of this work was carried out.

This research has made use of the NASA/IPAC Extragalactic Database
(NED) which is operated by the Jet Propulsion Laboratory, California
Institute of Technology, under contract with the National Aeronautics
and Space Administration.

\section*{References}


\reference Bardelli, S., Zucca, E., Malizia, A., Zamorani, G.,
Scaramella, R., Vettolani, G., 1996, A\&A, 305, 435

\reference Bardelli, S., Zucca, E., Zamorani, G., Vettolani, G.,
Scaramella, R., 1998, MNRAS, 296, 599

\reference Bardelli, S., Zucca, E., Zamorani, G., 1999, in
Observational Cosmology: The Development of Galaxy systems', Sesto
1998, (astro-ph/9811015)

\reference Drinkwater, M.J., Currie, M.J., Young, C.K., Hardy, E., Yearsley, J.M., 1996, MNRAS, 279, 595

\reference Infante L., Slezak E., Quintana H., 1996, A\&A 315, 657

\reference Kraan-Korteweg, R.C., Woudt, P.A., Cayatte, V., Fairall, A.P.,
Balkowski, C., Henning, P.A., 1996, Nature, 379, 519

\reference Kurtz M.J., Mink D.J., 1998, PASP, in press

\reference Metcalfe, N., Shanks, T., Fong, R., Jones, L.R., 1991, MNRAS, 249, 498

\reference Parker, Q.A., Watson, F.G., 1995, in Wide Field
Spectroscopy and the Distant Universe, 35th Herstmonceux Conference,
ed. S.J. Maddox, \& A. Aragon-Salamanca, (Singapore: World
Scientific), 33

\reference Parker, Q.A., 1997, in Wide Field Spectroscopy, 2nd
conference of the working group of IAU Commission 9 on Wide Field
Imaging, ed. Kontizas et al., (Dordrecht: Kluwer), 25

\reference Pierre, M., Bohringer, H., Ebeling, H., Voges, W.,
Schuecker, P., Cruddace, R., MacGillivray, H., 1994, A\&A 290, 725

\reference Tody, D. 1993, in Astronomical Data Analysis Software and
Systems II, A.S.P. Conference Ser., Vol 52, eds. R.J. Hanisch,
R.J.V. Brissenden, \& J. Barnes, 173.

\reference Tonry J., Davis M., 1979, AJ, 84, 1511.

\reference Quintana, H., Ramirez, A., Melnick, J., Raychaudhury, S.,
Slezak, E., 1995, AJ 110, 463 (Paper I)

\reference Quintana, H, Melnick, J., Proust, D., Infante, L., 1997,
A\&A Sup, 125, 247 (Paper II)

\reference Raychaudhury, S., 1989, Nature, 342, 251

\reference Scaramella R., Baiesi-Pillastrini, G, Chincarini, G.,
Vettolani, G., Zamorani, G., 1989, Nature 338, 562

 
\begin{table}
\caption{Abell Clusters in the Shapley Region}
\label{tab_clus}
\bigskip

{\small

\begin{center}

\begin{tabular}{crrrr}
\hline
 RA (B1950) Dec    & $cz$ & Name & Dist & Ref \\
                   & (\kms) \\  
\hline \\
12:47:12 $-$36:29:00 &     - & 3527   & 6 &  N    \\
12:51:18 $-$26:07:00 &     - & 1633   & 6 &  N    \\
12:51:36 $-$28:45:00 & 15854 & 3528   & 4 &  I    \\
12:52:54 $-$30:05:00 & 15960 & 3530   & 4 &  I    \\
12:54:24 $-$32:39:00 & 22454 & 3531   & 5 &  II   \\
12:54:36 $-$30:06:00 & 16100 & 3532   & 4 &  I    \\
12:56:18 $-$26:21:00 & 22994 & 1648   & - &  N    \\
12:57:00 $-$33:24:00 & 14330 & S718   & 4 &  II   \\
12:58:18 $-$32:10:00 &  5007 & 3537   & - &  N    \\
13:05:54 $-$34:18:00 & 27551 & 3542.1 & 4 &  II   \\
13:05:54 $-$34:18:00 & 39393 & 3542.2 & 4 &  II   \\
13:08:36 $-$33:49:00 & 22245 & 3545   & 6 &  II   \\ 
13:11:36 $-$29:11:00 & 22634 & 3549   & 5 &  II   \\
13:12:24 $-$33:23:00 & 17688 & S726   & 5 &  I    \\
13:16:06 $-$31:33:00 & 14818 & 3552   & 6 &  I    \\
13:16:24 $-$36:55:00 & 15110 & 3553   & 4 &  I    \\
13:16:42 $-$33:13:00 & 14570 & 3554   & 4 &  II   \\
13:18:00 $-$28:43:00 & 14810 & 3555   & 4 &  I    \\
13:18:42 $-$35:32:00 & 14960 & S729   & 5 &  I    \\
13:20:12 $-$34:37:00 & 15140 & S731   & 4 &  II   \\
13:21:18 $-$31:24:00 & 14130 & 3556.1 & 4 &  B    \\
13:21:18 $-$31:24:00 & 15066 & 3556.2 & 4 &  B    \\
13:22:06 $-$28:37:00 & 14270 & 3557.1 & 5 &  II   \\
13:22:06 $-$28:37:00 & 23084 & 3557.2 & 5 &  II   \\
13:23:00 $-$36:59:00 & 20806 & S733   & - &  N    \\
13:24:06 $-$26:51:00 & 10368 & 1736.1 & 3 &  I    \\
13:24:06 $-$26:51:00 & 13357 & 1736.2 & 3 &  I    \\
13:25:06 $-$31:14:00 & 14403 & 3558   & 3 &  B    \\
13:26:58 $-$31:21:00 & 14841 & 1327   & - &  N    \\
13:27:06 $-$29:16:00 & 13812 & 3559   & 4 &  I    \\
13:28:38 $-$31:33:43 & 12868 & 1329   & - &  I    \\
13:29:00 $-$32:58:00 & 13864 & 3560   & 3 &  I    \\
13:30:42 $-$31:25:00 & 14492 & 3562   & 3 &  B    \\
13:31:30 $-$34:58:00 & 14930 & 3564   & 3 &  I    \\
13:33:48 $-$33:43:00 &  3268 & 3565   & - &  N    \\
13:36:06 $-$35:18:00 & 15469 & 3566   & 4 &  II   \\ 
13:39:24 $-$26:01:00 & 32048 & 1771   & - &  N    \\
13:43:54 $-$37:40:00 & 11152 & 3570   & - &  N    \\
13:44:36 $-$32:37:00 & 11730 & 3571   & 2 &  I    \\
13:45:18 $-$33:08:00 & 12142 & 3572   & - &  N    \\
13:46:18 $-$30:03:00 &  4227 & 3574   & - &  N    \\
13:49:42 $-$32:38:00 & 11242 & 3575   & 4 &  II   \\
13:51:30 $-$27:36:00 & 14870 & 3577   & 4 &  II   \\
13:54:42 $-$24:29:00 & 11152 & 3578   & 3 &  II   \\
\hline
\end{tabular}
\end{center}
}

References: 
I: Quintana et al.\ (1995) (Paper I)
II: Quintana et al.\ (1997) (Paper II)
N: NASA/IPAC Extragalactic Database (NED)
B: Bardelli et al.\ (1998)

\end{table}

\begin{table}
\caption{Galaxies observed}
\label{tab_data}
\bigskip
\centering
{
\footnotesize
\begin{tabular}{rcrrr}
\hline
 RA (B1950) Dec    &field &    $m_R$ & $V$ & $\sigma_V$ \\
                   &      &     & \multicolumn{2}{c}{(\kms)} \\  
\hline
13:00:14.3 $-$36:21:34 & 382 & 15.86 &  3443 & 36 \\
13:00:35.5 $-$33:42:37 & 382 & 15.53 & 23713 & 68 \\
13:01:07.2 $-$36:16:08 & 382 & 14.32 & 10298 & 56 \\
13:01:28.5 $-$36:11:57 & 382 & 15.74 & 27469 & 33 \\
13:01:30.0 $-$33:36:27 & 382 & 15.80 & 21437 &126 \\
13:01:34.5 $-$37:14:51 & 382 & 15.61 & 15818 &101 \\
13:02:11.4 $-$36:34:14 & 382 & 15.84 & 17698 & 13 \\
13:02:38.8 $-$36:15:49 & 382 & 14.90 & 37525 & 71 \\
13:02:50.6 $-$34:01:55 & 382 & 15.86 & 25647 & 42 \\
13:03:01.9 $-$36:28:58 & 382 & 14.39 & 10092 & 72 \\
13:03:14.3 $-$37:20:00 & 382 & 15.50 & 14899 & 97 \\
13:03:16.0 $-$35:44:30 & 382 & 15.70 & 24066 & 96 \\
13:03:28.1 $-$36:44:17 & 382 & 15.65 & 14943 & 69 \\
13:03:44.6 $-$35:40:23 & 382 & 15.78 & 14737 & 72 \\
13:03:48.0 $-$33:55:36 & 382 & 15.62 & 25475 &184 \\
13:04:03.2 $-$36:42:55 & 382 & 15.92 & 15011 &215 \\
13:04:08.4 $-$37:06:37 & 382 & 15.65 & 14058 & 47 \\
13:04:32.6 $-$35:51:35 & 382 & 14.53 & 15099 & 75 \\
13:04:46.9 $-$32:34:17 & 382 & 14.24 & 15419 &132 \\
13:04:49.9 $-$34:09:57 & 382 & 15.97 & 18872 & 37 \\
13:04:56.2 $-$37:19:11 & 382 & 15.31 & 15024 & 60 \\
13:05:12.2 $-$37:06:25 & 382 & 15.48 & 13228 &117 \\
13:05:19.9 $-$33:00:31 & 382 & 15.64 & 10171 &150 \\
13:05:20.7 $-$34:03:32 & 382 & 15.69 & 15174 &163 \\
13:05:23.1 $-$35:56:31 & 382 & 15.87 & 15763 &139 \\
13:05:40.8 $-$34:09:29 & 382 & 15.19 & 15067 & 78 \\
13:06:10.3 $-$37:15:19 & 382 & 15.90 & 14312 & 81 \\
13:06:34.7 $-$34:29:38 & 382 & 15.21 & 13201 & 76 \\
13:06:47.3 $-$34:25:01 & 382 & 15.91 & 18578 & 77 \\
13:06:55.9 $-$35:20:35 & 382 & 15.40 & 28284 & 52 \\
13:07:07.2 $-$36:52:20 & 382 & 15.96 & 14350 & 61 \\
13:07:29.6 $-$32:36:43 & 382 & 14.51 &  9507 & 81 \\
13:07:34.8 $-$34:20:49 & 382 & 15.77 & 27522 & 60 \\
13:07:48.3 $-$37:02:06 & 382 & 14.28 & 14555 & 42 \\
13:08:01.8 $-$33:42:02 & 382 & 15.72 & 15113 & 95 \\
13:08:19.8 $-$32:43:36 & 382 & 15.87 & 43801 & 44 \\
13:08:20.1 $-$32:49:07 & 382 & 14.39 & 15899 & 84 \\
13:08:25.4 $-$36:24:37 & 382 & 15.11 &  3363 & 89 \\
13:08:27.2 $-$33:49:39 & 382 & 15.72 & 14837 & 25 \\
13:09:01.9 $-$34:02:24 & 382 & 15.82 & 27235 & 66 \\
13:09:07.1 $-$35:47:53 & 382 & 14.24 & 10491 & 54 \\
13:09:15.4 $-$37:11:40 & 382 & 15.59 & 13069 &119 \\
13:09:19.3 $-$34:19:50 & 382 & 14.92 &  3196 & 80 \\
13:09:34.3 $-$35:04:41 & 382 & 15.83 & 23360 & 77 \\
13:09:35.1 $-$37:04:57 & 382 & 15.96 & 10414 & 81 \\
13:09:39.0 $-$33:45:40 & 382 & 15.55 & 13063 &141 \\
13:09:53.9 $-$33:06:23 & 382 & 15.75 & 23141 & 79 \\
13:09:55.1 $-$35:59:01 & 382 & 15.04 & 14386 & 63 \\
13:10:13.6 $-$34:30:54 & 382 & 15.60 & 15159 &154 \\
13:10:31.9 $-$32:40:51 & 382 & 15.75 & 29991 & 72 \\
13:10:47.9 $-$33:06:26 & 382 & 15.99 & 29645 &117 \\
\hline
\end{tabular}}
{
\footnotesize
\begin{tabular}{rcrrr}
\hline
 RA (B1950) Dec    &field &    $m_R$ & $V$ & $\sigma_V$ \\
                   &      &     & \multicolumn{2}{c}{(\kms)} \\  
\hline
13:10:54.7 $-$35:10:08 & 382 & 14.99 &  7527 & 96 \\
13:11:02.1 $-$36:16:46 & 382 & 14.52 & 14367 & 76 \\
13:11:09.8 $-$33:39:12 & 382 & 14.21 & 15299 & 49 \\
13:11:22.4 $-$33:49:50 & 382 & 14.63 & 15355 & 81 \\
13:11:23.5 $-$36:57:36 & 382 & 15.21 & 10229 & 83 \\
13:11:30.6 $-$36:55:53 & 382 & 14.26 & 10688 & 59 \\
13:11:42.3 $-$33:07:11 & 382 & 14.46 &  8889 & 69 \\
13:11:54.3 $-$33:05:46 & 382 & 15.58 & 14649 &115 \\
13:12:05.7 $-$33:29:36 & 382 & 15.99 & 30594 &115 \\
13:12:19.3 $-$32:37:16 & 382 & 15.51 & 14280 & 90 \\
13:12:23.9 $-$33:22:29 & 382 & 14.20 & 14531 & 79 \\
13:12:39.3 $-$36:55:29 & 382 & 15.66 & 10048 & 85 \\
13:12:39.9 $-$32:26:47 & 444 & 14.15 & 13869 & 96 \\
13:13:13.3 $-$36:43:06 & 382 & 15.90 & 31971 & 88 \\
13:13:21.9 $-$32:37:24 & 382 & 14.71 & 15284 & 88 \\
13:13:34.6 $-$35:00:24 & 382 & 15.81 & 11473 & 98 \\
13:13:37.9 $-$32:35:56 & 382 & 15.92 & 14428 & 86 \\
13:13:40.5 $-$37:12:41 & 382 & 15.45 & 14637 & 23 \\
13:13:54.9 $-$32:59:57 & 382 & 15.08 & 14315 & 98 \\
13:13:59.8 $-$35:31:47 & 382 & 15.80 & 32384 &102 \\
13:14:30.4 $-$31:33:07 & 444 & 15.97 & 14998 &126 \\
13:14:35.9 $-$33:11:02 & 382 & 14.26 & 15334 & 48 \\
13:14:37.9 $-$33:39:08 & 382 & 14.86 & 15224 & 73 \\
13:14:44.4 $-$31:20:39 & 444 & 15.83 & 15632 &176 \\
13:15:00.6 $-$32:57:45 & 382 & 15.93 & 25689 & 64 \\
13:15:08.7 $-$37:10:05 & 382 & 14.68 &  7330 & 66 \\
13:15:17.0 $-$35:32:43 & 382 & 14.98 &  3265 & 94 \\
13:15:18.1 $-$29:53:31 & 444 & 15.34 &  9769 & 73 \\
13:15:18.6 $-$36:43:14 & 382 & 15.30 & 14921 & 28 \\
13:15:21.3 $-$31:45:59 & 444 & 14.64 &  4358 &114 \\
13:15:29.4 $-$37:02:54 & 382 & 15.06 & 15014 & 63 \\
13:15:38.5 $-$33:02:17 & 382 & 14.33 &  4342 & 51 \\
13:16:06.6 $-$33:09:45 & 382 & 15.55 & 13407 & 47 \\
13:16:07.8 $-$31:33:17 & 444 & 14.20 & 15641 &103 \\
13:16:17.3 $-$36:58:20 & 382 & 14.91 & 14311 & 86 \\
13:16:31.5 $-$33:02:46 & 382 & 15.99 & 15047 & 69 \\
13:16:38.9 $-$33:05:31 & 382 & 15.94 & 14960 & 70 \\
13:16:39.5 $-$34:50:23 & 382 & 14.48 &  1247 & 28 \\
13:16:40.8 $-$33:15:18 & 382 & 14.33 & 14981 & 98 \\
13:16:41.2 $-$33:39:11 & 382 & 15.17 &  8432 & 37 \\
13:16:49.5 $-$36:42:34 & 382 & 15.76 & 11687 & 66 \\
13:17:08.9 $-$34:02:03 & 382 & 15.43 & 23525 &110 \\
13:17:24.0 $-$34:50:33 & 382 & 15.64 & 13924 &121 \\
13:17:28.8 $-$36:49:52 & 382 & 15.16 &  2359 & 47 \\
13:17:38.4 $-$34:04:01 & 382 & 15.70 & 15106 &145 \\
13:17:47.5 $-$33:52:14 & 382 & 15.72 & 23532 & 69 \\
13:17:53.0 $-$34:40:14 & 382 & 15.73 & 15393 & 86 \\
13:18:05.5 $-$37:00:34 & 382 & 15.90 & 15980 & 78 \\
13:18:06.6 $-$27:47:20 & 444 & 15.97 & 15326 &122 \\
13:18:15.6 $-$32:47:18 & 382 & 15.23 & 15768 & 64 \\
13:18:16.1 $-$35:20:17 & 382 & 15.47 & 15654 & 85 \\
\hline
\end{tabular}
}
\end{table}

\begin{table}
\caption{Galaxies observed (continued)}
\bigskip
\centering
{
\footnotesize
\begin{tabular}{rcrrr}
\hline
 RA (B1950) Dec    &field &    $m_R$ & $V$ & $\sigma_V$ \\
                   &      &     & \multicolumn{2}{c}{(\kms)} \\  
\hline
13:18:26.7 $-$34:55:52 & 382 & 15.19 & 20402 & 58 \\
13:18:31.3 $-$35:33:05 & 382 & 15.42 & 15373 &163 \\
13:18:36.3 $-$32:30:51 & 444 & 14.59 & 14469 &110 \\
13:18:38.8 $-$35:54:22 & 382 & 15.02 & 15345 & 94 \\
13:18:50.7 $-$34:40:34 & 382 & 15.19 & 15222 &196 \\
13:18:58.0 $-$35:16:37 & 382 & 14.53 & 13903 &170 \\
13:19:01.6 $-$33:36:49 & 382 & 15.83 & 35049 &132 \\
13:19:07.6 $-$35:29:36 & 382 & 15.62 & 16535 & 91 \\
13:19:15.1 $-$28:06:35 & 444 & 14.58 & 14030 & 85 \\
13:19:21.1 $-$34:32:11 & 382 & 15.54 & 15664 &126 \\
13:19:26.8 $-$34:21:31 & 382 & 15.59 & 15667 &127 \\
13:19:39.1 $-$33:06:29 & 382 & 15.88 &  3497 & 96 \\
13:19:47.9 $-$35:36:32 & 382 & 15.50 & 14747 &142 \\
13:19:54.1 $-$33:10:27 & 382 & 14.26 & 14379 & 44 \\
13:20:04.8 $-$34:47:15 & 382 & 14.97 &  7237 & 96 \\
13:20:10.1 $-$32:27:07 & 382 & 14.78 &  8503 & 67 \\
13:20:13.1 $-$34:43:27 & 382 & 15.00 & 16498 & 84 \\
13:20:24.0 $-$35:03:57 & 382 & 15.76 & 14710 & 59 \\
13:20:52.7 $-$35:38:04 & 382 & 14.80 &  3792 & 78 \\
13:20:53.3 $-$34:11:58 & 382 & 15.45 & 15426 &126 \\
13:20:57.0 $-$31:26:49 & 444 & 15.92 & 15170 & 63 \\
13:20:58.5 $-$27:43:52 & 444 & 14.54 & 13674 & 69 \\
13:21:04.1 $-$34:24:00 & 382 & 14.49 & 14472 &103 \\
13:21:05.8 $-$31:55:18 & 444 & 14.91 & 15249 &107 \\
13:21:08.5 $-$29:02:44 & 444 & 15.63 & 14096 & 98 \\
13:21:30.5 $-$31:23:03 & 444 & 14.66 & 14390 &176 \\
13:21:33.1 $-$32:37:59 & 382 & 14.36 & 14251 &135 \\
13:21:37.9 $-$29:59:26 & 444 & 14.43 &  3634 & 98 \\
13:21:39.8 $-$33:42:58 & 382 & 15.35 & 14827 & 87 \\
13:21:51.8 $-$36:02:13 & 382 & 15.14 & 10771 & 96 \\
13:21:58.5 $-$34:06:56 & 382 & 15.04 &  8371 &106 \\
13:22:02.2 $-$31:18:01 & 444 & 16.00 & 14773 & 14 \\
13:22:20.4 $-$33:39:05 & 382 & 15.94 & 27194 &132 \\
13:22:46.5 $-$36:19:44 & 382 & 15.34 & 15208 & 42 \\
13:22:56.5 $-$37:01:27 & 382 & 15.44 & 10225 &123 \\
13:23:22.9 $-$36:47:09 & 382 & 15.98 & 19958 &150 \\
13:23:57.5 $-$31:16:09 & 444 & 16.00 & 14660 &117 \\
13:24:04.7 $-$33:55:06 & 382 & 15.89 & 14909 & 34 \\
13:24:08.5 $-$31:56:19 & 444 & 15.71 & 13839 & 84 \\
13:24:12.5 $-$29:48:56 & 444 & 14.88 &  1866 & 78 \\
13:24:15.0 $-$30:57:58 & 444 & 14.92 & 16332 &122 \\
13:24:18.1 $-$36:49:27 & 382 & 15.72 & 14367 & 76 \\
13:24:21.4 $-$37:00:34 & 382 & 15.62 & 14520 & 36 \\
13:24:28.2 $-$36:49:27 & 383 & 15.74 & 15098 &102 \\
13:24:28.4 $-$36:14:05 & 383 & 15.31 & 14658 & 62 \\
13:24:35.3 $-$32:02:27 & 444 & 15.48 & 14176 & 82 \\
13:24:58.9 $-$32:21:44 & 383 & 15.66 & 11832 & 16 \\
13:25:13.1 $-$31:30:38 & 444 & 15.97 & 15590 & 96 \\
13:25:14.4 $-$32:30:11 & 383 & 15.86 & 11585 & 57 \\
13:25:20.1 $-$31:25:21 & 444 & 15.17 & 12200 &135 \\
13:25:25.0 $-$29:59:01 & 444 & 15.17 & 12776 & 81 \\
\hline
\end{tabular}}
{
\footnotesize
\begin{tabular}{rcrrr}
\hline
 RA (B1950) Dec    &field &    $m_R$ & $V$ & $\sigma_V$ \\
                   &      &     & \multicolumn{2}{c}{(\kms)} \\  
\hline
13:25:41.9 $-$33:44:20 & 383 & 15.86 & 24911 &114 \\
13:25:42.1 $-$33:56:52 & 383 & 15.88 & 14661 &129 \\
13:26:14.4 $-$28:34:52 & 444 & 14.20 & 12289 & 58 \\
13:26:27.7 $-$31:47:04 & 444 & 14.74 & 14060 &114 \\
13:26:39.2 $-$31:17:35 & 444 & 14.03 & 15430 & 87 \\
13:26:52.9 $-$28:00:28 & 444 & 14.44 & 10015 &119 \\
13:27:09.0 $-$28:59:02 & 444 & 15.83 & 14201 &114 \\
13:27:14.2 $-$27:43:01 & 444 & 15.12 & 10237 &114 \\
13:27:22.6 $-$32:32:55 & 383 & 14.33 & 15666 & 78 \\
13:27:24.1 $-$29:14:38 & 444 & 15.23 & 14488 & 30 \\
13:27:39.0 $-$34:21:48 & 383 & 15.95 & 21467 & 84 \\
13:27:51.3 $-$31:36:28 & 444 & 15.96 & 14470 & 75 \\
13:28:11.4 $-$32:21:51 & 383 & 15.68 & 15986 & 64 \\
13:28:19.5 $-$32:28:46 & 383 & 15.73 &  3468 & 95 \\
13:28:32.5 $-$33:38:26 & 383 & 15.58 & 22089 & 90 \\
13:28:58.4 $-$36:41:55 & 383 & 15.35 & 14996 & 97 \\
13:29:07.6 $-$36:03:10 & 383 & 15.33 & 20314 & 76 \\
13:29:14.5 $-$33:31:57 & 383 & 14.93 &  8770 & 66 \\
13:29:16.4 $-$33:03:38 & 383 & 15.08 & 13640 & 98 \\
13:29:24.5 $-$33:07:15 & 383 & 14.71 & 15243 & 81 \\
13:29:33.4 $-$32:58:55 & 383 & 14.95 & 13916 & 47 \\
13:29:33.7 $-$37:15:24 & 383 & 14.96 & 15295 & 29 \\
13:29:40.1 $-$33:54:18 & 383 & 15.87 & 14770 & 96 \\
13:29:40.9 $-$37:02:12 & 383 & 15.58 & 15687 &101 \\
13:29:49.3 $-$34:53:51 & 383 & 15.23 & 15122 & 39 \\
13:30:00.5 $-$34:27:42 & 383 & 15.47 & 36879 & 77 \\
13:30:00.7 $-$28:45:40 & 444 & 15.57 &  5526 & 93 \\
13:30:04.4 $-$32:54:49 & 383 & 15.19 & 15698 & 50 \\
13:30:10.9 $-$32:50:38 & 383 & 14.62 & 14951 & 53 \\
13:30:13.4 $-$37:14:36 & 383 & 15.79 & 15164 & 67 \\
13:30:14.9 $-$33:02:41 & 383 & 15.65 & 14838 &115 \\
13:30:15.5 $-$34:11:36 & 383 & 14.84 &  7491 & 16 \\
13:30:19.9 $-$31:20:49 & 444 & 15.96 & 14559 &123 \\
13:30:34.1 $-$33:41:34 & 383 & 15.99 & 22010 &143 \\
13:30:45.7 $-$31:13:03 & 444 & 14.97 & 15201 & 74 \\
13:30:53.3 $-$35:08:58 & 383 & 15.92 & 15377 &147 \\
13:31:01.2 $-$35:07:39 & 383 & 15.88 & 21640 & 89 \\
13:31:17.6 $-$31:13:59 & 444 & 15.86 & 14842 & 89 \\
13:31:17.8 $-$27:54:11 & 444 & 14.85 &  9187 & 74 \\
13:31:19.5 $-$34:48:41 & 383 & 15.28 & 14917 & 69 \\
13:31:21.8 $-$32:43:17 & 383 & 14.43 & 15177 & 70 \\
13:31:25.2 $-$29:07:15 & 444 & 15.35 & 14008 & 96 \\
13:31:30.3 $-$35:09:30 & 383 & 15.37 &  9416 & 65 \\
13:31:33.2 $-$31:28:46 & 444 & 14.89 & 16271 & 85 \\
13:31:39.7 $-$36:33:12 & 383 & 16.00 & 15274 &150 \\
13:31:48.9 $-$34:46:49 & 383 & 15.55 &  2367 & 21 \\
13:31:53.3 $-$34:58:33 & 383 & 15.35 & 16579 & 20 \\
13:32:17.4 $-$33:46:46 & 383 & 14.98 &  3878 & 52 \\
13:32:28.5 $-$31:10:15 & 444 & 14.86 & 14845 & 86 \\
13:32:31.3 $-$33:54:20 & 383 & 14.18 &  7208 & 66 \\
13:32:48.9 $-$31:27:42 & 444 & 15.94 & 11794 & 98 \\
\hline
\end{tabular}
}

\end{table}

\begin{table}
\caption{Galaxies observed (continued)}
\bigskip
\centering
{
\footnotesize
\begin{tabular}{rcrrr}
\hline
 RA (B1950) Dec    &field &    $m_R$ & $V$ & $\sigma_V$ \\
                   &      &     & \multicolumn{2}{c}{(\kms)} \\  
\hline
13:32:54.0 $-$35:18:55 & 383 & 14.42 & 15438 & 54 \\
13:33:21.5 $-$36:38:27 & 383 & 15.17 & 15135 & 51 \\
13:33:37.8 $-$32:50:15 & 383 & 15.46 & 15688 &105 \\
13:33:41.2 $-$33:49:11 & 383 & 14.19 &  3787 & 43 \\
13:33:45.9 $-$28:50:04 & 444 & 14.16 &  4516 & 89 \\
13:34:05.6 $-$34:08:42 & 383 & 15.14 &  4133 & 76 \\
13:34:11.1 $-$33:34:08 & 383 & 14.29 & 13931 & 87 \\
13:34:12.6 $-$28:16:45 & 444 & 15.90 & 15262 & 98 \\
13:34:28.4 $-$34:38:09 & 383 & 15.50 & 29247 &103 \\
13:34:43.7 $-$31:37:28 & 444 & 15.38 & 11598 & 98 \\
13:34:45.1 $-$34:16:42 & 383 & 15.67 & 22108 & 68 \\
13:34:49.6 $-$33:35:55 & 383 & 15.54 & 11289 & 85 \\
13:34:50.2 $-$35:14:29 & 383 & 16.00 & 15162 & 89 \\
13:34:59.3 $-$35:23:12 & 383 & 14.68 & 15695 & 71 \\
13:35:15.2 $-$34:48:29 & 383 & 15.47 & 15162 & 92 \\
13:35:21.3 $-$31:41:48 & 444 & 15.26 & 14694 &127 \\
13:35:27.8 $-$35:42:28 & 383 & 15.50 & 11327 & 74 \\
13:35:47.0 $-$34:17:07 & 383 & 15.24 & 15420 &116 \\
13:35:52.1 $-$35:13:11 & 383 & 15.91 & 14685 & 37 \\
13:36:00.6 $-$36:11:38 & 383 & 14.66 & 13789 & 55 \\
13:36:04.4 $-$35:22:16 & 383 & 15.50 & 15223 & 89 \\
13:36:04.7 $-$36:20:26 & 383 & 15.80 & 13755 & 74 \\
13:36:08.6 $-$32:20:51 & 444 & 14.76 & 16298 & 77 \\
13:36:15.9 $-$36:39:42 & 383 & 15.30 & 15741 & 44 \\
13:36:18.3 $-$33:47:22 & 383 & 15.75 & 21245 & 28 \\
13:36:32.6 $-$35:12:49 & 383 & 15.47 & 19147 & 68 \\
13:36:43.3 $-$35:24:54 & 383 & 15.19 & 15791 & 78 \\
13:36:44.2 $-$33:45:54 & 383 & 15.13 & 15268 & 76 \\
13:36:49.8 $-$35:41:31 & 383 & 15.49 & 29346 & 77 \\
13:36:56.1 $-$33:24:07 & 383 & 14.54 & 15381 & 43 \\
13:36:56.6 $-$32:49:51 & 383 & 15.43 & 15345 &135 \\
13:37:12.1 $-$32:48:53 & 383 & 15.43 & 11841 & 59 \\
13:37:14.8 $-$32:35:15 & 383 & 14.56 &  7254 & 14 \\
13:37:15.1 $-$34:05:41 & 383 & 15.66 & 21441 & 89 \\
13:37:37.8 $-$33:44:08 & 383 & 15.91 & 14867 & 61 \\
13:37:39.9 $-$35:25:16 & 383 & 15.54 & 15641 & 57 \\
13:37:59.8 $-$33:53:02 & 383 & 15.99 & 15579 & 30 \\
13:38:01.9 $-$33:25:24 & 383 & 15.47 & 15167 & 66 \\
13:38:12.4 $-$34:07:13 & 383 & 15.57 & 15485 & 98 \\
13:38:22.3 $-$35:24:09 & 383 & 14.38 & 15068 & 85 \\
13:38:39.8 $-$34:07:45 & 383 & 15.56 & 16973 & 66 \\
13:38:40.8 $-$33:41:14 & 383 & 15.39 & 15161 & 48 \\
13:38:49.2 $-$33:49:06 & 383 & 15.76 & 14610 & 47 \\
13:38:52.4 $-$35:37:34 & 383 & 14.88 & 11378 & 62 \\
13:38:57.2 $-$32:55:58 & 383 & 15.49 & 11945 & 65 \\
13:39:27.0 $-$36:56:20 & 383 & 14.84 & 11288 & 31 \\
13:39:35.6 $-$34:37:32 & 383 & 15.61 &  4454 & 51 \\
13:39:43.7 $-$37:21:55 & 383 & 15.78 & 10122 & 76 \\
13:39:55.3 $-$35:55:15 & 383 & 15.64 & 21887 &155 \\
13:40:02.0 $-$34:43:11 & 383 & 15.07 & 16170 & 50 \\
13:40:26.8 $-$36:04:17 & 383 & 15.18 & 24883 &122 \\
\hline
\end{tabular}}
{
\footnotesize
\begin{tabular}{rcrrr}
\hline
 RA (B1950) Dec    &field &    $m_R$ & $V$ & $\sigma_V$ \\
                   &      &     & \multicolumn{2}{c}{(\kms)} \\  
\hline
13:40:33.4 $-$32:59:43 & 383 & 15.55 & 12267 & 97 \\
13:40:34.0 $-$37:23:43 & 383 & 15.64 & 11113 & 52 \\
13:40:40.2 $-$34:31:47 & 383 & 15.08 & 17030 & 72 \\
13:40:42.0 $-$36:09:58 & 383 & 14.60 &  4311 & 59 \\
13:40:52.6 $-$34:49:27 & 383 & 14.32 & 16227 &183 \\
13:40:59.8 $-$35:31:12 & 383 & 14.38 & 11637 & 28 \\
13:41:14.7 $-$34:05:19 & 383 & 14.87 & 11202 & 76 \\
13:41:23.2 $-$36:22:42 & 383 & 15.81 & 11498 & 46 \\
13:41:23.9 $-$33:51:28 & 383 & 15.21 & 11818 & 71 \\
13:41:47.3 $-$32:24:12 & 383 & 15.53 & 12438 & 76 \\
13:41:49.3 $-$35:11:00 & 383 & 15.45 & 37694 & 72 \\
13:41:52.3 $-$33:21:05 & 383 & 14.32 & 11063 & 31 \\
13:41:55.5 $-$34:22:45 & 383 & 15.24 & 14660 & 48 \\
13:42:08.6 $-$34:58:58 & 383 & 15.67 & 17045 & 66 \\
13:42:10.1 $-$32:38:25 & 383 & 15.77 & 12085 & 83 \\
13:42:25.1 $-$32:23:47 & 383 & 14.23 &  9469 & 41 \\
13:42:33.6 $-$36:31:58 & 383 & 15.05 & 11531 & 64 \\
13:42:38.9 $-$35:01:09 & 383 & 15.79 & 17237 & 50 \\
13:42:56.1 $-$33:46:54 & 383 & 15.10 & 15058 & 33 \\
13:43:08.2 $-$33:29:39 & 383 & 14.32 & 11662 & 44 \\
13:43:17.6 $-$34:14:50 & 383 & 15.67 & 24806 & 35 \\
13:43:29.2 $-$32:55:23 & 383 & 15.54 & 11760 & 52 \\
13:43:32.6 $-$35:50:50 & 383 & 15.11 & 11548 & 27 \\
13:43:33.0 $-$32:24:59 & 383 & 15.36 & 12884 & 52 \\
13:44:02.0 $-$33:04:05 & 383 & 15.98 & 11113 &157 \\
13:44:06.5 $-$37:12:33 & 383 & 15.71 & 11357 & 50 \\
13:44:06.7 $-$35:03:53 & 383 & 15.23 & 11453 & 47 \\
13:44:15.6 $-$32:39:29 & 383 & 15.76 & 12304 &159 \\
13:44:16.1 $-$33:08:38 & 383 & 15.44 & 11518 & 77 \\
13:44:19.6 $-$33:23:43 & 383 & 15.87 & 13010 & 70 \\
13:44:22.8 $-$36:07:31 & 383 & 15.04 & 10963 & 66 \\
13:44:29.3 $-$33:21:14 & 383 & 15.38 & 16117 & 48 \\
13:44:29.5 $-$32:50:50 & 383 & 14.99 & 11694 & 49 \\
13:44:56.8 $-$32:56:05 & 383 & 14.92 & 10328 & 29 \\
13:45:05.5 $-$32:37:12 & 383 & 15.59 & 11789 & 80 \\
13:45:11.0 $-$33:06:07 & 383 & 14.35 & 11901 & 34 \\
13:45:18.8 $-$32:32:24 & 383 & 15.74 & 12585 &138 \\
13:45:29.9 $-$34:26:11 & 383 & 15.15 & 13896 & 88 \\
13:45:37.7 $-$35:26:14 & 383 & 15.90 & 30107 & 78 \\
13:45:47.9 $-$34:24:28 & 383 & 15.29 & 27740 & 73 \\
13:45:49.1 $-$37:14:52 & 383 & 15.47 & 11128 & 83 \\
13:46:04.9 $-$32:36:32 & 383 & 14.39 & 11603 & 49 \\
13:46:05.4 $-$33:09:44 & 383 & 14.72 & 10868 & 31 \\
13:46:06.3 $-$33:43:18 & 383 & 15.06 & 11327 & 35 \\
13:46:56.0 $-$35:14:07 & 383 & 15.60 & 28576 &122 \\
13:46:59.9 $-$33:02:31 & 383 & 15.78 & 16129 & 67 \\
13:47:03.7 $-$36:30:57 & 383 & 15.22 & 30091 & 89 \\
13:47:22.5 $-$32:49:37 & 383 & 14.35 & 10522 & 50 \\
13:47:38.1 $-$32:25:54 & 383 & 15.49 & 11098 & 51 \\
13:47:48.9 $-$32:38:51 & 383 & 15.12 & 11270 & 37 \\
13:48:08.4 $-$35:50:11 & 383 & 15.46 & 22610 & 95 \\
\hline
\end{tabular}
}

\end{table} 

\end{document}